\pdfoutput=1
\documentclass[aps, reprint, superscriptaddress, letterpaper, floatfix, prb]{revtex4-1}
\usepackage{graphicx}
\usepackage{upgreek}
\usepackage{color}
\usepackage{subcaption}
\usepackage{tabularx}
\usepackage{dcolumn}
\usepackage{bm}
\usepackage{amsmath}
\usepackage{scrextend}
\usepackage[normalem]{ulem}
\usepackage{hyperref}
\usepackage{cleveref}
\crefname{figure}{Figure}{Figures}
\Crefname{figure}{Figure}{Figures}
\crefname{table}{Table}{Tables}
\Crefname{table}{Table}{Tables}

\begin{document}

\title{The magnetic Lyddane-Sachs-Teller relation}

\author{Viktor Rindert}
\email[Electronic mail: ]{viktor.rindert@ftf.lth.se}
\affiliation{NanoLund and Solid State Physics, Lund University, S-22100 Lund, Sweden}
\author{Vanya~Darakchieva}
\affiliation{NanoLund and Solid State Physics, Lund University, S-22100 Lund, Sweden}
\affiliation{Department of Physics, Chemistry, and Biology (IFM), Link{\"o}ping University, SE 58183, Link{\"o}ping, Sweden}
\author{Tapati Sarkar}
\affiliation{Department of Materials Science and Engineering, Uppsala University, Box 35, SE-751 03 Uppsala, Sweden.}
\author{Mathias Schubert}
\affiliation{Department of Electrical and Computer Engineering, University of Nebraska-Lincoln, Lincoln, NE 68588, USA}
\affiliation{NanoLund and Solid State Physics, Lund University, S-22100 Lund, Sweden}

\date{\today}

\begin{abstract}
We describe a magnetic relation in analogy to the well-known dielectric Lyddane-Sachs-Teller relation [R. H. Lyddane, R. G. Sachs, E. Teller, Phys. Rev. 59, 673 (1941)]. This magnetic relation follows directly from the model equations for nuclear induction due to fast oscillating electromagnetic fields [F. Bloch, Phys. Rev. 70, 460 (1946)] and relates the static permeability with the product over all ratios of antiresonance and resonance frequencies associated with all magnetic excitations within a given specimen. The magnetic relation differs significantly from its dielectric analog where the static properties are related to ratios of the squares of resonance frequencies. We demonstrate the validity of the magnetic Lyddane-Sachs-Teller relation using optical magnetization data from terahertz electron magnetic resonance spectroscopic ellipsometry measurements on an iron-doped semiconductor crystal of gallium nitride. 
\end{abstract}
\maketitle

\noindent \textbf{Introduction:}
We describe here a fundamental law in magnetism that relates static properties with optical phenomena in magnetism. This law -- which we will term here the magnetic Lyddane-Sachs-Teller relation -- has a well-known analog – the dielectric Lyddane-Sachs-Teller (LST) relation.\cite{1941LST} The latter was described by Nobel Laureate Edward Teller and coworkers within a seminal paper in 1941. This fundamental law has since entered textbooks in solid state physics.\cite{Klingshirn} We derive the magnetic relation here from another, equally important contribution in physics, the nuclear induction paper by Nobel Laureate Felix Bloch on the interaction of nuclear magnetic moments with fast oscillating electromagnetic fields.\cite{Bloch_orig} This 1946 paper led to the discovery of magnetic resonance, which today is perhaps one of the most widespread techniques used in science. The magnetic relation derived here is similar yet differs significantly from its dielectric analog as will be discussed further below. We demonstrate here the validity of the magnetic Lyddane-Sachs-Teller relation using optical magnetization data available from terahertz spectroscopic ellipsometry measurements on an iron-doped semiconductor crystal of gallium nitride. We developed instruments recently to measure magnetic resonance as predicted by Felix Bloch continuously as a function of frequency and with complete polarization information.\cite{SiCpaper} Analysis of poles and zeros of the real and imaginary parts of the permeability function provides access to magnetic resonance and antiresonance frequencies analogous to the transverse and longitudinal optical frequencies in the dielectric function across dielectric lattice mode excitations. Thus we can also demonstrate here experimental evidence for the correctness of the magnetic Lyddane-Sachs-Teller relation.

The LST relation establishes a fundamental equality between two critical ratios in materials exhibiting optical lattice vibrations.\cite{1941LST} Specifically, the relation equates the square of the ratio of frequencies between longitudinal optic (LO; $\omega_{\mathrm{LO}}$) and transverse optic (TO; $\omega_{\mathrm{TO}}$) lattice vibrations at long wavelengths, for ionic crystals with one optical phonon branch, with the ratio of the static ($\varepsilon_{\mathrm{dc}}$) and high-frequency dielectric permeability ($\varepsilon_{\infty}$)
\begin{equation}
\frac{\varepsilon_{\mathrm{dc}}}{\varepsilon_{\infty}}=\frac{\omega^2_{\mathrm{LO}}}{\omega^2_{\mathrm{TO}}}.
\end{equation}
Transverse resonance occurs at very large displacement and longitudinal resonance occurs when the field-induced polarization compensates its vacuum contribution (antiresonance). For materials with multiple ($N$) optical phonon branches, extensions to the LST relation have been made by Barker\cite{Barker} and Berreman and Unterwald\cite{BU}
\begin{equation}
\frac{\upepsilon_{\mathrm{dc}}}{\upepsilon_{\infty}}=\prod^{N}_{l=1}\frac{\omega^2_{\mathrm{LO,l}}}{\omega^2_{\mathrm{TO,l}}}.
\label{eq:LST}
\end{equation}
Further generalizations have been proposed by Cochran and Cowley,\cite{1962COCHRAN} who have extended the theory to include the phonon displacement vector. This version is applicable when the principal polarization axes align with orthogonal directions. A coordinate invariant LST relation was introduced, which holds for all crystal symmetries, including monoclininc and triclinic.\cite{2016SchubertInvariant} The coordinate invariant LST was derived from the microscopic Born-Huang description of polar lattice vibrations, and relates the determinant of the permittivity tensor at zero and infinite frequencies to the LO and TO resonance modes defined thereby\cite{2016SchubertInvariant}
\begin{equation}
    \det (\boldsymbol{\upepsilon}^{-1}(\omega_\textrm{TO}))=0, \textbf{   } \ \ \det (\boldsymbol{\upepsilon}(\omega_\textrm{LO})) = 0,
\end{equation}
\begin{equation}
\frac{\det (\boldsymbol{\upepsilon}(\omega=0))}{\det (\boldsymbol{\upepsilon}_{\infty})}=\prod^{N}_{l=1}\frac{\omega^2_{\mathrm{LO,l}}}{\omega^2_{\mathrm{TO,l}}}.
\label{eq:CILST}
\end{equation}
This generalization has proved pivotal in analyzing phonon modes in monoclinic crystal structures, advancing our understanding of material properties.\cite{2019SchubertPhonon,2019SchubertLPP,2018MockPhonon, Mock2017, Stokey}

In this letter, we extend the concept of the LST relation towards magnetic dipole transitions. We use equations introduced by Felix Bloch for the frequency response of magnetic moments in slowly varying external fields.\cite{Bloch_orig} Hence, in the linear optics regime, we must find an expression for the permeability tensor $\boldsymbol{\upmu}$ which relates the magnetization $\boldsymbol{M}$ to the external magnetic field $\boldsymbol{H}$ via $\boldsymbol{M}= \boldsymbol{\upmu} (\omega) \boldsymbol{H}$.

Spectroscopic information about the response function tensor is needed to make use of the LST relation. Such spectroscopic information about magnetic resonances can be gained using THz Mueller matrix ellipsometry, as demonstrated recently for the N defect in SiC.\cite{SiCpaper} Different model approaches can be employed to obtain frequencies, amplitudes, transverse relaxation times, and spin volume density parameters.\cite{SiCpaper, RichterPRBFebGO2024, rindert2024bloch} Here, we present a method in which the permeability tensor derived from the Bloch equations permits to determine the dc magnetization from the analysis of TO and LO modes associated with magnetic resonance(s). We apply this theoretical framework to experimental data obtained using our in-house built THz electron magnetic resonance (EPR) ellipsometer and validate our results with results from dc magnetization measurements using a superconducting quantum interference device (SQUID) magnetometer, demonstrating very good agreement. We make use of the well-known magnetic properties of Fe-doped Gallium Nitride (GaN) as an example.

\noindent \textbf{Theory:} The Bloch equations describe the time-evolution of a magnetic moment vector $\boldsymbol{M}$ and can be written as\cite{Bloch_orig,boltonchapter10, BARANOV1997611,rindert2024bloch}
\begin{align}
    \begin{split}
    &\frac{\partial \boldsymbol{M}}{\partial t} = \gamma_e ( \boldsymbol{M} \times \boldsymbol{H} ) - \frac{\boldsymbol{M}(\hat{e}_x + \hat{e}_y)}{T_2} -  \frac{(\boldsymbol{M} - \boldsymbol{M}_0)\hat{e}_z}{T_1},
    \end{split}
    \label{eq:bloch1}
\end{align}
where $\times$ denotes the cross product. Here, $\gamma_e$ represents the electron gyromagnetic ratio and is taken to be positive valued. The unit vectors in the laboratory coordinate system, $(\hat{e}_x,\hat{e}_y,\hat{e}_z)$, correspond to the \textit{x}, \textit{y}, and \textit{z} axes, respectively, with the \textit{z}-axis aligned parallel to the direction of the magnetic field. Additionally, $T_1$ and $T_2$ are the phenomenological longitudinal and transverse relaxation times, respectively. These relaxation times describe how the system returns to equilibrium when conditions, e.g., the magnetic field direction change, and contribute a wealth of information regarding the electronic structure and molecular dynamics.\cite{Eaton2000} The longitudinal relaxation time generally depends on spin-lattice interactions, and the transverse relaxation time depends on spin-spin interactions.\cite{Abragam_Bleaney_1970} 

The total magnetic field is given by
\begin{equation}
\boldsymbol{H} = H_0 \hat{e}_z + H_x(t) \hat{e}_x + H_y(t) \hat{e}_y
\label{eq:bfield}
\end{equation}
where $H_0$ represents the static component of the magnetic field aligned with the \textit{z}-axis, and $H_x(t)$ and $H_y(t)$ denote the time-varying components of the magnetic field in the \textit{x} and \textit{y} directions, respectively. These latter components are attributed to the rotating magnetic field generated by polarized light traversing the system. Upon substituting \eqref{eq:bfield} into the Bloch equations, we derive the following system of equations:
\begin{align}
\begin{split}
& \frac{\partial M_x}{\partial t} = \omega_0 M_y - \frac{M_x}{T_2} + \gamma_e M_z H_y(t),\\
& \frac{\partial M_y}{\partial t} = -\omega_0 M_x - \frac{M_y}{T_2} - \gamma_e M_z H_x(t),\\
& \frac{\partial M_z}{\partial t} = - \frac{M_z - M_0}{T_1} + \gamma_e (M_y H_x(t) - M_x H_y(t)),
\end{split}
\label{eq:Bloch_Bfield}
\end{align}
where we have introduced $\omega_0=\gamma_eH_0$, which corresponds to the angular frequency of the magnetization's precession around the \textit{z}-axis. When the magnetic field components $H_x(t)$ and $H_y(t)$ are significantly smaller than $H_0$, a condition described as the low-power limit, the magnetization along the \textit{z}-axis, $M_z$, can be approximated as remaining at its equilibrium value, $M_0$, and ${\partial M_z}/{\partial t} = 0$. This simplification implies a negligible influence of saturation effects on the system.\cite{Slichter1990} We can, in the low-power limit, further simplify by the substitution $\gamma_eM_z=\gamma_eM_0= \gamma_eH_0\chi_0=\omega_0\chi_0$ where $\chi_0\equiv M_0/H_0$ is the dc magnetic susceptibility, and take the time derivative of \eqref{eq:Bloch_Bfield}
\begin{align}
    \begin{split}
       & \frac{\partial M_x^2}{\partial t^2} =  \omega_0 \frac{\partial M_y}{\partial t}- \frac{1}{T_2}\frac{\partial M_x}{\partial t} + \chi_0 \omega_0 \frac{\partial H_y(t)}{\partial t} \\
       & \frac{\partial M_y^2}{\partial t^2} =  -\omega_0 \frac{\partial M_x}{\partial t}- \frac{1}{T_2}\frac{\partial M_y}{\partial t} - \chi_0 \omega_0 \frac{\partial H_x(t)}{\partial t} 
    \end{split}
    \label{Bloch3}
\end{align}
such that we can decouple the \textit{x}- and \textit{y}-direction by inserting \eqref{eq:Bloch_Bfield} into \eqref{Bloch3}, which results in
\begin{align}
    \begin{split}
        & \frac{\partial M_x^2}{\partial t^2} + \frac{1}{T_2}\frac{\partial M_x}{\partial t} = + \chi_0 \omega_0 \frac{\partial H_y}{\partial t} \\& -\omega_0 ( \omega_0 M_x - \chi_0 \omega_0 H_x  - \frac{M_y}{T_2}), \\ & \frac{\partial M_y^2}{\partial t^2} + \frac{1}{T_2}\frac{\partial M_y}{\partial t} = - \chi_0 \omega_0 \frac{\partial H_x}{\partial t} \\ &-
        \omega_0 ( \omega_0 M_y - \chi_0 \omega_0 H_y  + \frac{M_x}{T_2}).
    \end{split}
\end{align}

Next we assume that $\omega_0 >> 1/T_2$ to discard $M_x/T_2$ and $M_y/T_2$, and move all $M_x$ and $M_y$ terms to one side
\begin{align}
    \begin{split}
        & \frac{\partial M_x^2}{\partial t^2} + \frac{1}{T_2} \frac{\partial M_x}{\partial t}+ \omega_0^2 M_x= \omega_0 (   \chi_0 \omega_0 H_x ) + \chi_0 \omega_0 \frac{\partial H_y(t)}{\partial t}, \\
        & \frac{\partial M_y^2}{\partial t^2} + \frac{1}{T_2}\frac{\partial M_y}{\partial t} +\omega_0^2 M_y = \omega_0 (  \chi_0 \omega_0 H_y) - \chi_0 \omega_0 \frac{\partial H_x(t)}{\partial t}. 
    \end{split}
    \label{eq:bloch4}
\end{align}
Finally, to get the frequency response, the Fourier-transform
\begin{equation}
    F(\omega) = \int_{-\infty}^\infty f(t) \exp(-i\omega t) \textrm{d}t,
\end{equation}
is used to transform \eqref{eq:bloch4} into
\begin{align}
    \begin{split}
       & M_x(\omega_0^2 - i\omega/T_2- \omega^2) = \omega_0^2 \chi_0 H_x -  i \chi_0 \omega_0 \omega H_y, \\
       & M_y(\omega_0^2 - i\omega/T_2- \omega^2) = \omega_0^2 \chi_0 H_y +  i \chi_0 \omega_0 \omega H_x,
    \end{split}
\end{align}
and in tensor form
\begin{align}
    \begin{split}
    &\begin{pmatrix}
        M_x \\ M_y
    \end{pmatrix} = \boldsymbol{\chi_M} \begin{pmatrix}
        H_x \\ H_y
    \end{pmatrix} =\\ &\chi_0 
    \begin{pmatrix}
        \frac{\omega_0^2 }{\omega_0^2 - i\omega /T_2- \omega^2 } & -i \frac{\omega \omega_0}{\omega_0^2 - i\omega /T_2- \omega^2 } \\
        i\frac{\omega  \omega_0}{\omega_0^2 - i\omega /T_2- \omega^2 } & \frac{\omega_0^2 }{\omega_0^2 - i\omega /T_2- \omega^2 }
    \end{pmatrix}\begin{pmatrix}
        H_x \\ H_y
    \end{pmatrix},
    \end{split}
    \label{eq:xhitensor-Polder}
\end{align}
where tensor $\mathbf{\chi_{M}}$ has a structure similar to the Polder tensor for the magnetic susceptibility of ferrites.\cite{doi:10.1080/14786444908561215}
To arrive at the LST relation and its generalizations, it is convenient to neglect the linewidth broadening ($1/T_2=0$), as it does not affect the final result. For the case of systems with spin $S>1/2$, it is assumed that each of the $2S$ transitions adheres to the same lineshape, and the Bloch permeability is written as
\begin{equation}
    \boldsymbol{\upmu} = \boldsymbol{I} + \sum_{j=1}^{2S}\begin{pmatrix}
        \frac{\chi_{0,j}\omega_{0,j}^2 }{\omega_{0,j}^2 - \omega^2 } & i \frac{\chi_{0,j}\omega\omega_{0,j} }{\omega_{0,j}^2 - \omega^2 } & 0\\
        -i\frac{\omega\omega_{0,j} }{\omega_{0,j}^2 - \omega^2 } & \frac{\chi_{0,j}\omega_{0,j}^2 }{\omega_{0,j}^2 - \omega^2 } & 0 \\ 0 & 0 & \chi_{0,j}
    \end{pmatrix},
    \label{eq:prbtensor_sum}
\end{equation}
where each of the 2S spin transitions contributes independently to the static susceptibility $\chi_0=\sum_{j=1}^{2S}\chi_{0,j}$, similar to the extension of the LST relation to multiple optical phonon branches.\cite{Barker,BU} This analogy invites a comparison between spin and phonon transitions, given the well-characterized nature of phonon modes in low-symmetry materials.\cite{2016SchubertInvariant, 2018MockPhonon} To begin with, setting $\omega=0$ in Eq.~\eqref{eq:prbtensor_sum} we recover the dc response $\boldsymbol{\upmu}(\omega=0) =\upmu_\mathrm{dc}\textbf{I} $,
where $\upmu_\mathrm{dc} = (1 + \chi_0)$ denotes the dc magnetic permeability. It is noteworthy that this model does not capture diamagnetic contributions. Magnetic dipole transitions outside the available spectral range such as radio-frequency induced nuclear magnetic resonance also do not contribute to the summation of $\chi_0$. We utilize the coordinate invariant LST and search for the eigenresonances
\begin{equation}
    \det (\boldsymbol{\upmu}^{-1}(\omega_\textrm{TO})) = 0, \quad \det (\boldsymbol{\upmu}(\omega_\textrm{LO})) = 0.
\end{equation}
It can be easily verified that the determinant of $\boldsymbol{\upmu}$ diverges when the frequency $\omega$ equals one of the resonant frequencies $\omega_{0,j}$. Thus, the TO modes correspond to the resonance frequencies, $\omega_{\textrm{TO},j} = \omega_{0,j}$. However, identifying the LO modes requires a more detailed analysis. The determinant is given by
\begin{widetext}
\begin{align}
\begin{split}
    \det (\boldsymbol{\upmu}) &= \upmu_\mathrm{dc} \left( 1 + 2\sum_{j=1}^{2S} \frac{\chi_{0,j} \omega_{\textrm{TO},j}^2}{\omega_{\textrm{TO},j}^2 - \omega^2} + \left(\sum_{j=1}^{2S} \frac{\chi_{0,j} \omega_{\textrm{TO},j}^2}{\omega_{\textrm{TO},j}^2 - \omega^2}\right)^2 \right. \left. - \ \omega^2 \left(\sum_{j=1}^{2S} \frac{\chi_{0,j} \omega_{\textrm{TO},j}}{\omega_{\textrm{TO},j}^2 - \omega^2}\right)^2 \right).
\end{split}
\end{align}
We make use of the form $a^2-b^2=(a+b)(a-b)$ and rewrite accordingly
\begin{align}
         \det (\boldsymbol{\upmu}) &
          =\upmu_\mathrm{dc}\left\{1 + \sum_{j=1}^{2S} \chi_{0,j}\frac{ \omega_{\textrm{TO},j}^2 - \omega\omega_{\textrm{TO},j}}{\omega_{\textrm{TO},j}^2 - \omega^2} \right\}\left[1 + \sum_{j=1}^{2S} \chi_{0,j}\frac{ \omega_{\textrm{TO},j}^2 + \omega\omega_{\textrm{TO},j}}{\omega_{\textrm{TO},j}^2 - \omega^2} \right], \\
         & = \upmu_\mathrm{dc}\left\{1 + \sum_{j=1}^{2S} \chi_{0,j}\frac{ \omega_{\textrm{TO},j}}{(\omega_{\textrm{TO},j} + \omega)} \right\}
         \left[1 + \sum_{j=1}^{2S} \chi_{0,j}\frac{ \omega_{\textrm{TO},j}}{(\omega_{\textrm{TO},j} - \omega)} \right],\\&=
         \upmu_\mathrm{dc}
         \left\{\frac{1}{\prod_{j=1}^{2S} \omega_{\textrm{TO},j}+\omega}P^{2S}(-\omega) \right\}
         \left[\frac{1}{\prod_{j=1}^{2S} \omega_{\textrm{TO},j}-\omega}P^{2S}(\omega) \right].
         \end{align}
\end{widetext}
Here, $P^{2S}(\omega)$ is a polynomial of degree 2S, which, according to the Fundamental Theorem of Algebra can be written as the product of its linear factors, where the roots correspond to the LO modes
\begin{align}
         \det (\boldsymbol{\upmu})&=\upmu_\mathrm{dc}
         \left\{ \prod_{j=1}^{2S}\frac{\omega_{\textrm{LO}, j}+\omega}{\omega_{\textrm{TO}, j} + \omega} \right\}
         \left[ \prod_{j=1}^{2S}\frac{\omega_{\textrm{LO}, j}-\omega}{\omega_{\textrm{TO}, j} - \omega} \right].
         \label{detfunction}
    \end{align}
Setting $\omega=0$ we obtain
\begin{equation}
    \det (\boldsymbol{\upmu}(\omega=0)) = \upmu_{\mathrm{dc}} \prod_{j=1}^{2S} \frac{\omega_{\textrm{LO},j}^2}{\omega_{\textrm{TO},j}^2} = \upmu_{\mathrm{dc}}^3,
    \label{eq:CI-pmLST}
\end{equation}
and after dividing by $\mu_{\mathrm{dc}}$ and taking the root,
\begin{equation}
    \upmu_\mathrm{dc} = \prod_{j=1}^{2S} \frac{\omega_{\textrm{LO}}}{\omega_{\textrm{TO}}}.
    \label{eq:pmLST}
\end{equation}
Equation~\ref{eq:pmLST} is the main finding of this letter -- the magnetic analog of the LST relation. Equation~\ref{eq:CI-pmLST} is independent on the choice of coordinates and thus establishes the coordinate invariant magnetic LST relation. Note the significant difference between the magnetic LST (Eq.~\ref{eq:pmLST}) and LST (Eq.~\ref{eq:LST}) relations, where the dielectric resonance frequencies appear squared in the latter while the magnetic resonance frequencies appear in first order in the former. We further propose to rewrite Eq.~\eqref{detfunction} and introduce a generalized magnetic permeability function akin to the Berreman-Unterwald form\cite{BU}
\begin{equation}
\det (\boldsymbol{\upmu}(\omega))= \upmu_\mathrm{dc} \prod_{j=1}^{2S} \frac{\omega_{\textrm{LO},j}^2-\omega^2}{\omega_{\textrm{TO},j}^2- \omega^2}.
\end{equation}
Here, relaxation times $T_{2,j}$ and $T^{\star}_{2,j}$ can be introduced
\begin{equation}
\det (\boldsymbol{\upmu}(\omega))= \upmu_\mathrm{dc} \prod_{j=1}^{2S} \frac{\omega_{\textrm{LO},j}^2-\omega^2-i\omega/T^{\star}_{2,j}}{\omega_{\textrm{TO},j}^2- \omega^2-i\omega/T_{2,j}},
\label{eq:mBUform}
\end{equation}
and we propose Eq.~\ref{eq:mBUform} to be used for analysis of measured magnetic permeability tensor spectra. Parameters $T_{2,j}, T^{\star}_{2,j}$ refer to the transverse relaxation times for resonances $\omega_{\mathrm{TO},j}$, $\omega_{\mathrm{LO},j}$, respectively. Such spectra maybe gained from investigation of polarized magnetic reststrahlen bands in ferromagnets, for example.\cite{magnonpolaritons_restrahlen, Magneticpolaritonmodes, Kullmann_1984} 
\begin{table}
\centering
\caption{Calculated parameters from a best-match model using non-linear least squares optimization on THz EPR ellipsometry data at 15~K and a magnetic field of $\pm4.42$~T. Uncertainties for the last significant digit are provided in parentheses. Uncertainties for parameters $\chi_{0,j}$ and $T_{2,j}$ were estimated using the square root of the covariance matrix derived from the optimization. The uncertainties for frequencies are assumed to match uncertainty in the magnetic field strength, known to five significant digits.}
\begin{tabular}{lccccc}
        \hline\hline
        $j$ & 1 & 2 & 3 & 4 & 5 \\
        \hline
        $\omega_{\textrm{TO},j}$\\ (GHz) & 121.87(1) & 122.65(1) & 124.07(1) & 125.72(1) & 127.22(1) \\
        $\omega_{\textrm{LO},j}$-$\omega_{\textrm{TO},j}$\\ (kHz) & 23.037(1) & 65.624(1) & 111.12(1) & 177.03(1) & 301.34(1) \\
        $\chi_{0,j}$\\ ($10^{-8}$) & $19(3)$ & $54(3)$ & $90(4)$ & $141(5)$ & $237(7)$ \\
        $T_{2,j}$\\ (ns) & $2.6(7)$ & $3.2(3)$ & $4.9(3)$ & $3.3(2)$ & $3.3(2)$ \\
        \hline\hline
\end{tabular}
\label{tab2}
\end{table}
We demonstrate the validity of the magnetic LST relation by performing spectroscopic THz EPR ellipsometry measurements and subsequent analysis using the permeability tensor in Eq.~\eqref{eq:prbtensor_sum} including the relaxation time parameters.
\begin{align}
    \begin{split}
    &\boldsymbol{\upmu} = \boldsymbol{I} + \\&\sum_{j=1}^{2S}\chi_{0,j}\begin{pmatrix}
        \frac{\omega_{0,j}^2 }{\omega_{0,j}^2 - \omega^2 -i\omega/T_{2,j}} & \frac{i\omega\omega_{0,j} }{\omega_{0,j}^2 - \omega^2 -i\omega/T_{2,j}} & 0\\
        \frac{-i\omega\omega_{0,j} }{\omega_{0,j}^2 - \omega^2 -i\omega/T_{2,j}} & \frac{\omega_{0,j}^2 }{\omega_{0,j}^2 - \omega^2 -i\omega/T_{2,j}} & 0 \\ 0 & 0 & 1
    \end{pmatrix}.
    \end{split}
    \label{eq:prbtensor_sum_broadened}
\end{align}
We obtain the dc magnetization by summation over all transition amplitudes and compare the results with SQUID magnetization measurements. The TO modes are then known and the LO modes are then found numerically by determining the roots of Eq.~\eqref{eq:prbtensor_sum}. The product of their ratios is then tested against the results obtained from SQUID analysis as well, i.e., the validity of Eq.~\ref{eq:pmLST} is then confirmed.

\begin{figure*}[!tbp]
    \centering
    \begin{minipage}[t]{0.48\linewidth}
        \centering
        \includegraphics[width=\linewidth]{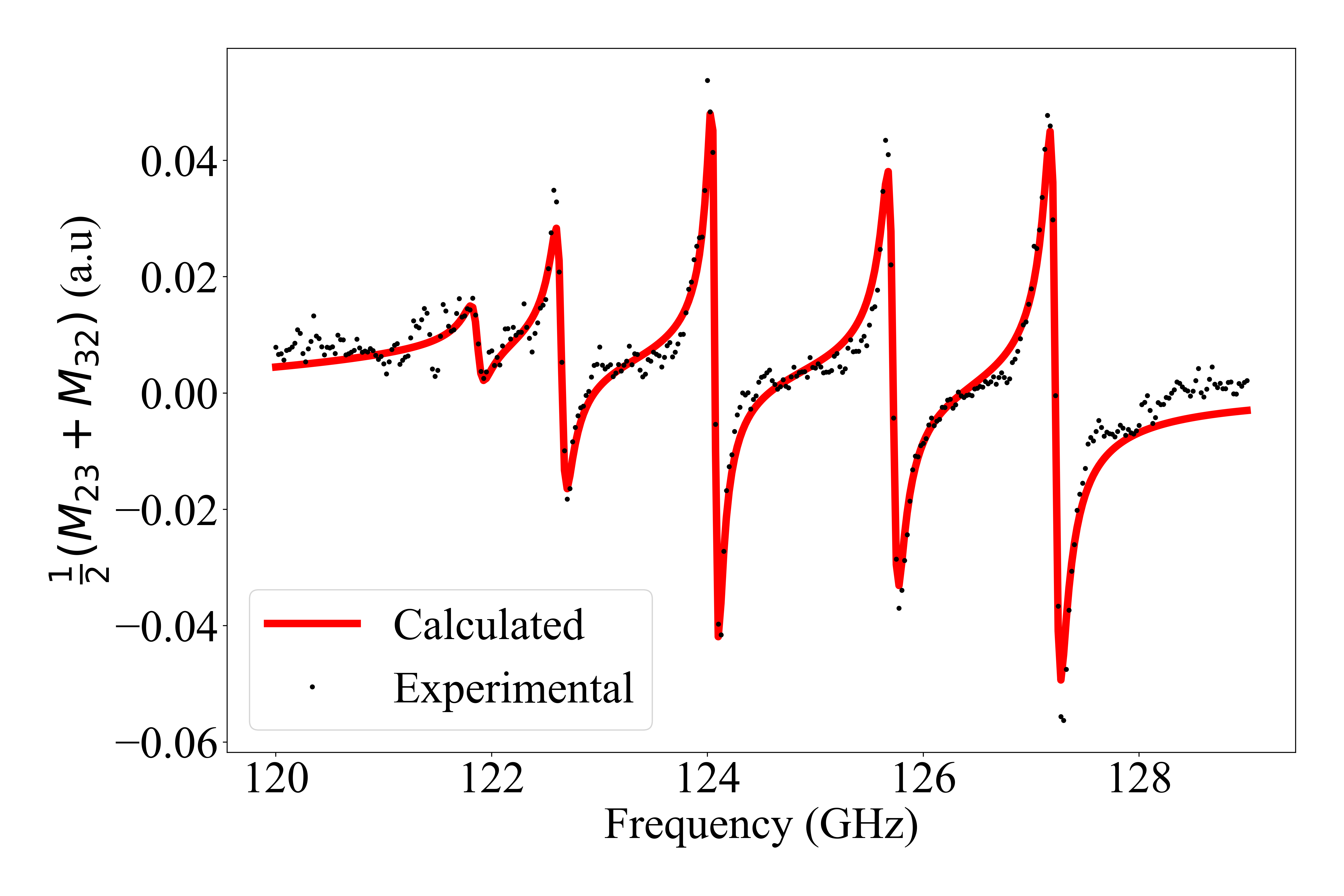}
        \caption{Experimental (black dots) and best-match model calculated (red line) data from a frequency-swept THz EPR Mueller matrix measurement, performed at 15~K and magnetic field $\pm4.42$~T. The sample is an Iron-doped GaN substrate.}
        \label{fig:res}
    \end{minipage}
    \hfill
    \begin{minipage}[t]{0.48\linewidth}
        \centering
        \includegraphics[width=\linewidth]{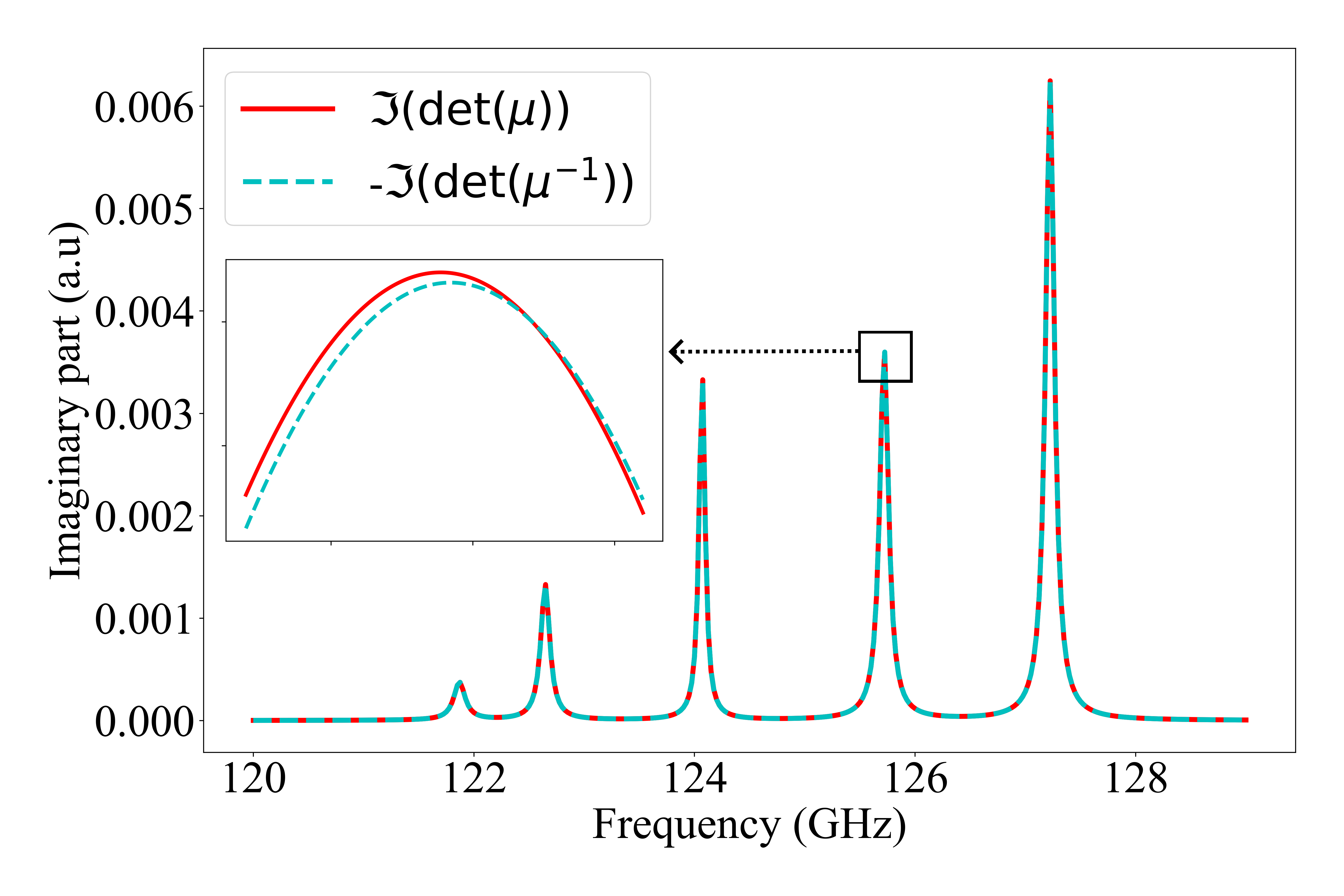}
        \caption{Imaginary part of the determinant of the frequency-dependent permeability tensor $\boldsymbol{\upmu}$ (red solid line) and the negative of the imaginary part of the determinant of the frequency-dependent inverse permeability tensor $\boldsymbol{\upmu^{-1}}$ (cyan dashed line). Resonance ($\omega_{\mathrm{TO},j}$) and antiresonance ($\omega_{\mathrm{LO},j}$) peak in $\Im\boldsymbol{\upmu}$ and $-\Im\boldsymbol{\upmu^{-1}}$, respectively. The inset enlarges the difference between transitions $j=4$.}
        \label{fig:inv_mu}
    \end{minipage}
\end{figure*}
\noindent \textbf{Method:} We conducted THz EPR ellipsometry and SQUID measurements on an Iron-doped GaN sample. The sample is a bulk single crystal with c-plane (0001) surface orientation produced via halide vapor phase epitaxy, measuring approximately 8~mm $\times$ 5~mm with a thickness of 0.989~mm (see supp. material).\cite{Paskova_APL2006}

For the THz EPR ellipsometry measurements, we used a superconducting split-coil magnet system\cite{2018LuEllipsometer} capable of generating magnetic fields from -8~T to 8~T with a field homogeneity of about 3000~ppm across the sample volume. The Mueller matrix elements were measured using a custom-built THz ellipsometer across a spectral range of 120~GHz to 129~GHz in 50~MHz increments, employing a tunable single-frequency continuous-wave source with a frequency bandwidth of approximately 50~kHz. A solid-state synthesizer frequency source, multiplied by a signal generator extension (Virginia Diodes, Inc.), provided precise digital control over frequency and duty cycle.

Intensity readings were collected at a Golay cell detector under various polarizer and analyzer configurations to obtain the top-left 3$\times$3 section of the Mueller matrix. Measurements were executed in a reflection setup with the sample positioned at a 45$^{\circ}$ angle of incidence between the split coils, aligning the magnetic field parallel to the incident beam. The magnetic field was oriented at 45$^{\circ}$ relative to the crystallographic $c$-axis of the Fe-doped GaN sample. By rotating the sample around its surface normal we aligned the two crystallographically equivalent gallium sites in the wurtzite lattice structure relative to the magnetic field. The alignment was done such that the zero-field splitting of the two different Ga site occupying Fe atoms resulted in equivalent spin levels thereby reducing the effect of multiplicity from the quintuplet spin transitions. Thus the two quintuplets coincide within the THz-EPR ellipsometry spectra.\cite{rindert2024bloch}

Measurements were conducted at a magnetic field strength of 4.42~T, repeated at -4.42~T, and also without a magnetic field to subtract background data, allowing for the extraction of small-signal difference data. The sample's temperature was maintained at 15~K throughout the measurement process. To model the response, we applied Eq.~\eqref{eq:prbtensor_sum_broadened} and the Berreman 4$\times$4 matrix formalism, and we compared the model calculated data with the experimental data by minimizing the difference using a least-squares method (see also supplementary material).

The dc magnetic susceptibility of the Fe-doped GaN single crystal was recorded as a function of temperature under a magnetic field of 4.42~T using a SQUID magnetometer from Quantum Design Inc. The sample was attached to a piece of paper using insulating varnish (GE Varnish, Oxford Instruments) and mounted with its plane parallel to the direction of the magnetic field. To eliminate the diamagnetic contribution, we made an additional magnetization versus magnetic field measurement. The diamagnetic susceptibility was then estimated from the slope of the high-field data and used to obtain the corrected value of the magnetization using the equation: $M_{\mathrm{corr}} = M_{\mathrm{exp}}-\chi_{\mathrm{dia}}H$, where $M_{\mathrm{corr}}$ is the corrected value of the magnetization, $M_{\mathrm{exp}}$ is the experimentally measured value, $\chi_{\mathrm{dia}}$ is the slope of the high-field region in the magnetization versus magnetic field curve, and $H$ is the magnetic field.

\noindent \textbf{Results and discussion:} The experimental results alongside the corresponding best-fit model are presented in Fig.~\hyperref[fig:res]{1}. A very good agreement between the best-match model calculated and experimental data is evident. This substantiates our theoretical approach in linking optical modes with static magnetic properties, as can be done for phononic resonances that follow a Lorentzian lineshape. In Fig.~\hyperref[fig:inv_mu]{2}, the imaginary parts of the calculated function $\det(\upmu)$ (Eq.~\ref{eq:prbtensor_sum_broadened}) and its inverse are plotted to illustrate the small differences between the associated TO and LO modes. For all transitions, differences are observed on the order of tens of kHz, and numerical values are tabulated in Tab.~\hyperref[tab2]{1}. By inserting the calculated optical modes into Eq.~\eqref{eq:pmLST}, we obtain the relative static permeability of $1+(5.4\pm0.2)\cdot10^{-6}$ within a 95\% confidence interval. We note that the largest source of uncertainty is the accuracy by which the sample thickness parameter is known (see also the supplementary material).

A volume magnetization of $M_{\textrm{Vol}} =$18.9~A/m was obtained at 15~K from the SQUID magnetometry measurement. The static permeability is then given by the relation
\begin{align}
\begin{split}
    &\upmu_\mathrm{dc} = 1 +\frac{M_{\textrm{Vol}}}{H} =\\&  1 + \frac{18.9 \textrm{ A/m}}{3.52\cdot10^6 \textrm{ A/m}} = 1 + 5.37 \cdot 10^{-6}.
    \end{split}
\end{align}
Given that the SQUID measurement was conducted with the static magnetic field aligned parallel to the sample surface, i.e., with $H$ parallel to the GaN $c$ axis, comparison with the THz EPR result requires extrapolating the relative static magnetic susceptibility for a scenario where the magnetic field is oriented at 45$^{\circ}$ relative to the GaN $c$ axis (see the supplementary material). After correction of the SQUID result for the magnetic field geometry a value of $1+5.32\cdot10^{-6}$ is obtained, in very good agreement with the ellipsometry result. Hence, we conclude an excellent match between the theory presented here and the results of the conducted experiments. We conclude the validity of the magnetic LST relation, therefore, for magnetic transitions such as those occurring at defects in semiconductor materials.

The resonant interaction of electromagnetic waves with intrinsic excitations in matter has traditionally been referred to as polariton coupling.\cite{1951Huang} Examples include phonon polaritons (coupling of infrared light with lattice excitations) or exciton polaritons (coupling of light with excitons).\cite{PEKAR195811} The characteristics of the coupling are often manifested by resonance and antiresonance conditions, for example, the transverse and longitudinal optical phonon modes leading to associated pairs of polariton branches.\cite{PhysRev.112.1555} The introduction of $\omega_{\mathrm{TO},j}$ and $\omega_{\mathrm{LO},j}$ for each EPR transition in this work calls for the investigation of the corresponding polariton properties and we propose to explore the associated magnetic polaritons for their potential use in future photonic applications.    

\section{Acknowledgments}
This work is supported by the Swedish Research Council under Grants No. 2016-00889 and No. 2022-04812, by the Knut and
Alice Wallenberg Foundation under award “Wide-bandgap semiconductors for next generation quantum components” (Grant No. 2018.0071), by the Swedish Foundation for Strategic Research under Grant No. EM16-0024, by the Swedish Governmental Agency
for Innovation Systems VINNOVA under the Competence Center Program Grant No. 2022-03139, and by the Swedish Government Strategic Research Area NanoLund and in Materials Science on Functional Materials at Link\"oping University, Faculty Grant SFO Mat LiU No. 009-00971. M.S. acknowledges support by the National Science Foundation under awards ECCS 2329940, and OIA-2044049 Emergent Quantum Materials and Technologies (EQUATE), by Air Force Office of Scientific Research under awards FA9550-19-S-0003, FA9550-21-1-0259, and FA9550-23-1-0574 DEF, and by the University of Nebraska Foundation. M.S. acknowledges support from the J.~A.~Woollam Foundation. T.S. gratefully acknowledges funding from the Swedish Research Council (grant number: 2021-03675).
\bibliography{refs}

\begin{thebibliography}{29}%
\makeatletter
\providecommand \@ifxundefined [1]{%
 \@ifx{#1\undefined}
}%
\providecommand \@ifnum [1]{%
 \ifnum #1\expandafter \@firstoftwo
 \else \expandafter \@secondoftwo
 \fi
}%
\providecommand \@ifx [1]{%
 \ifx #1\expandafter \@firstoftwo
 \else \expandafter \@secondoftwo
 \fi
}%
\providecommand \natexlab [1]{#1}%
\providecommand \enquote  [1]{``#1''}%
\providecommand \bibnamefont  [1]{#1}%
\providecommand \bibfnamefont [1]{#1}%
\providecommand \citenamefont [1]{#1}%
\providecommand \href@noop [0]{\@secondoftwo}%
\providecommand \href [0]{\begingroup \@sanitize@url \@href}%
\providecommand \@href[1]{\@@startlink{#1}\@@href}%
\providecommand \@@href[1]{\endgroup#1\@@endlink}%
\providecommand \@sanitize@url [0]{\catcode `\\12\catcode `\$12\catcode `\&12\catcode `\#12\catcode `\^12\catcode `\_12\catcode `\%12\relax}%
\providecommand \@@startlink[1]{}%
\providecommand \@@endlink[0]{}%
\providecommand \url  [0]{\begingroup\@sanitize@url \@url }%
\providecommand \@url [1]{\endgroup\@href {#1}{\urlprefix }}%
\providecommand \urlprefix  [0]{URL }%
\providecommand \Eprint [0]{\href }%
\providecommand \doibase [0]{http://dx.doi.org/}%
\providecommand \selectlanguage [0]{\@gobble}%
\providecommand \bibinfo  [0]{\@secondoftwo}%
\providecommand \bibfield  [0]{\@secondoftwo}%
\providecommand \translation [1]{[#1]}%
\providecommand \BibitemOpen [0]{}%
\providecommand \bibitemStop [0]{}%
\providecommand \bibitemNoStop [0]{.\EOS\space}%
\providecommand \EOS [0]{\spacefactor3000\relax}%
\providecommand \BibitemShut  [1]{\csname bibitem#1\endcsname}%
\let\auto@bib@innerbib\@empty
\bibitem [{\citenamefont {Lyddane}\ \emph {et~al.}(1941)\citenamefont {Lyddane}, \citenamefont {Sachs},\ and\ \citenamefont {Teller}}]{1941LST}%
  \BibitemOpen
  \bibfield  {author} {\bibinfo {author} {\bibfnamefont {R.~H.}\ \bibnamefont {Lyddane}}, \bibinfo {author} {\bibfnamefont {R.~G.}\ \bibnamefont {Sachs}}, \ and\ \bibinfo {author} {\bibfnamefont {E.}~\bibnamefont {Teller}},\ }\href {\doibase 10.1103/PhysRev.59.673} {\bibfield  {journal} {\bibinfo  {journal} {Phys. Rev.}\ }\textbf {\bibinfo {volume} {59}},\ \bibinfo {pages} {673} (\bibinfo {year} {1941})}\BibitemShut {NoStop}%
\bibitem [{\citenamefont {Klingshirn}(2012)}]{Klingshirn}%
  \BibitemOpen
  \bibfield  {author} {\bibinfo {author} {\bibfnamefont {C.~F.}\ \bibnamefont {Klingshirn}},\ }\href@noop {} {\emph {\bibinfo {title} {Semiconductor Optics}}}\ (\bibinfo  {publisher} {Springer Berlin},\ \bibinfo {year} {2012})\BibitemShut {NoStop}%
\bibitem [{\citenamefont {Bloch}(1946)}]{Bloch_orig}%
  \BibitemOpen
  \bibfield  {author} {\bibinfo {author} {\bibfnamefont {F.}~\bibnamefont {Bloch}},\ }\href {\doibase 10.1103/PhysRev.70.460} {\bibfield  {journal} {\bibinfo  {journal} {Phys. Rev.}\ }\textbf {\bibinfo {volume} {70}},\ \bibinfo {pages} {460} (\bibinfo {year} {1946})}\BibitemShut {NoStop}%
\bibitem [{\citenamefont {Schubert}\ \emph {et~al.}(2022)\citenamefont {Schubert}, \citenamefont {Knight}, \citenamefont {Richter}, \citenamefont {Kühne}, \citenamefont {Stanishev}, \citenamefont {Ruder}, \citenamefont {Stokey}, \citenamefont {Korlacki}, \citenamefont {Irmscher}, \citenamefont {Neugebauer},\ and\ \citenamefont {Darakchieva}}]{SiCpaper}%
  \BibitemOpen
  \bibfield  {author} {\bibinfo {author} {\bibfnamefont {M.}~\bibnamefont {Schubert}}, \bibinfo {author} {\bibfnamefont {S.}~\bibnamefont {Knight}}, \bibinfo {author} {\bibfnamefont {S.}~\bibnamefont {Richter}}, \bibinfo {author} {\bibfnamefont {P.}~\bibnamefont {Kühne}}, \bibinfo {author} {\bibfnamefont {V.}~\bibnamefont {Stanishev}}, \bibinfo {author} {\bibfnamefont {A.}~\bibnamefont {Ruder}}, \bibinfo {author} {\bibfnamefont {M.}~\bibnamefont {Stokey}}, \bibinfo {author} {\bibfnamefont {R.}~\bibnamefont {Korlacki}}, \bibinfo {author} {\bibfnamefont {K.}~\bibnamefont {Irmscher}}, \bibinfo {author} {\bibfnamefont {P.}~\bibnamefont {Neugebauer}}, \ and\ \bibinfo {author} {\bibfnamefont {V.}~\bibnamefont {Darakchieva}},\ }\href {\doibase 10.1063/5.0082353} {\bibfield  {journal} {\bibinfo  {journal} {Appl. Phys. Lett}\ }\textbf {\bibinfo {volume} {120}},\ \bibinfo {pages} {102101} (\bibinfo {year} {2022})},\ \Eprint {http://arxiv.org/abs/https://doi.org/10.1063/5.0082353} {https://doi.org/10.1063/5.0082353}
  \BibitemShut {NoStop}%
\bibitem [{\citenamefont {Barker}(1964)}]{Barker}%
  \BibitemOpen
  \bibfield  {author} {\bibinfo {author} {\bibfnamefont {A.~S.}\ \bibnamefont {Barker}},\ }\href {\doibase 10.1103/PhysRev.136.A1290} {\bibfield  {journal} {\bibinfo  {journal} {Phys. Rev.}\ }\textbf {\bibinfo {volume} {136}},\ \bibinfo {pages} {A1290} (\bibinfo {year} {1964})}\BibitemShut {NoStop}%
\bibitem [{\citenamefont {Berreman}\ and\ \citenamefont {Unterwald}(1968)}]{BU}%
  \BibitemOpen
  \bibfield  {author} {\bibinfo {author} {\bibfnamefont {D.~W.}\ \bibnamefont {Berreman}}\ and\ \bibinfo {author} {\bibfnamefont {F.~C.}\ \bibnamefont {Unterwald}},\ }\href {\doibase 10.1103/PhysRev.174.791} {\bibfield  {journal} {\bibinfo  {journal} {Phys. Rev.}\ }\textbf {\bibinfo {volume} {174}},\ \bibinfo {pages} {791} (\bibinfo {year} {1968})}\BibitemShut {NoStop}%
\bibitem [{\citenamefont {Cochran}\ and\ \citenamefont {Cowley}(1962)}]{1962COCHRAN}%
  \BibitemOpen
  \bibfield  {author} {\bibinfo {author} {\bibfnamefont {W.}~\bibnamefont {Cochran}}\ and\ \bibinfo {author} {\bibfnamefont {R.}~\bibnamefont {Cowley}},\ }\href {\doibase https://doi.org/10.1016/0022-3697(62)90084-7} {\bibfield  {journal} {\bibinfo  {journal} {Journal of Physics and Chemistry of Solids}\ }\textbf {\bibinfo {volume} {23}},\ \bibinfo {pages} {447} (\bibinfo {year} {1962})}\BibitemShut {NoStop}%
\bibitem [{\citenamefont {Schubert}(2016)}]{2016SchubertInvariant}%
  \BibitemOpen
  \bibfield  {author} {\bibinfo {author} {\bibfnamefont {M.}~\bibnamefont {Schubert}},\ }\href {\doibase 10.1103/PhysRevLett.117.215502} {\bibfield  {journal} {\bibinfo  {journal} {Phys. Rev. Lett.}\ }\textbf {\bibinfo {volume} {117}},\ \bibinfo {pages} {215502} (\bibinfo {year} {2016})}\BibitemShut {NoStop}%
\bibitem [{\citenamefont {Schubert}\ \emph {et~al.}(2019{\natexlab{a}})\citenamefont {Schubert}, \citenamefont {Mock}, \citenamefont {Korlacki},\ and\ \citenamefont {Darakchieva}}]{2019SchubertPhonon}%
  \BibitemOpen
  \bibfield  {author} {\bibinfo {author} {\bibfnamefont {M.}~\bibnamefont {Schubert}}, \bibinfo {author} {\bibfnamefont {A.}~\bibnamefont {Mock}}, \bibinfo {author} {\bibfnamefont {R.}~\bibnamefont {Korlacki}}, \ and\ \bibinfo {author} {\bibfnamefont {V.}~\bibnamefont {Darakchieva}},\ }\href {\doibase 10.1103/PhysRevB.99.041201} {\bibfield  {journal} {\bibinfo  {journal} {Phys. Rev. B}\ }\textbf {\bibinfo {volume} {99}},\ \bibinfo {pages} {041201} (\bibinfo {year} {2019}{\natexlab{a}})}\BibitemShut {NoStop}%
\bibitem [{\citenamefont {Schubert}\ \emph {et~al.}(2019{\natexlab{b}})\citenamefont {Schubert}, \citenamefont {Mock}, \citenamefont {Korlacki}, \citenamefont {Knight}, \citenamefont {Galazka}, \citenamefont {Wagner}, \citenamefont {Wheeler}, \citenamefont {Tadjer}, \citenamefont {Goto},\ and\ \citenamefont {Darakchieva}}]{2019SchubertLPP}%
  \BibitemOpen
  \bibfield  {author} {\bibinfo {author} {\bibfnamefont {M.}~\bibnamefont {Schubert}}, \bibinfo {author} {\bibfnamefont {A.}~\bibnamefont {Mock}}, \bibinfo {author} {\bibfnamefont {R.}~\bibnamefont {Korlacki}}, \bibinfo {author} {\bibfnamefont {S.}~\bibnamefont {Knight}}, \bibinfo {author} {\bibfnamefont {Z.}~\bibnamefont {Galazka}}, \bibinfo {author} {\bibfnamefont {G.}~\bibnamefont {Wagner}}, \bibinfo {author} {\bibfnamefont {V.}~\bibnamefont {Wheeler}}, \bibinfo {author} {\bibfnamefont {M.}~\bibnamefont {Tadjer}}, \bibinfo {author} {\bibfnamefont {K.}~\bibnamefont {Goto}}, \ and\ \bibinfo {author} {\bibfnamefont {V.}~\bibnamefont {Darakchieva}},\ }\href {\doibase 10.1063/1.5089145} {\bibfield  {journal} {\bibinfo  {journal} {Applied Physics Letters}\ }\textbf {\bibinfo {volume} {114}},\ \bibinfo {pages} {102102} (\bibinfo {year} {2019}{\natexlab{b}})},\ \Eprint {http://arxiv.org/abs/https://pubs.aip.org/aip/apl/article-pdf/doi/10.1063/1.5089145/13584189/102102\_1\_online.pdf}
  {https://pubs.aip.org/aip/apl/article-pdf/doi/10.1063/1.5089145/13584189/102102\_1\_online.pdf} \BibitemShut {NoStop}%
\bibitem [{\citenamefont {Mock}\ \emph {et~al.}(2018)\citenamefont {Mock}, \citenamefont {Korlacki}, \citenamefont {Knight},\ and\ \citenamefont {Schubert}}]{2018MockPhonon}%
  \BibitemOpen
  \bibfield  {author} {\bibinfo {author} {\bibfnamefont {A.}~\bibnamefont {Mock}}, \bibinfo {author} {\bibfnamefont {R.}~\bibnamefont {Korlacki}}, \bibinfo {author} {\bibfnamefont {S.}~\bibnamefont {Knight}}, \ and\ \bibinfo {author} {\bibfnamefont {M.}~\bibnamefont {Schubert}},\ }\href {\doibase 10.1103/PhysRevB.97.165203} {\bibfield  {journal} {\bibinfo  {journal} {Phys. Rev. B}\ }\textbf {\bibinfo {volume} {97}},\ \bibinfo {pages} {165203} (\bibinfo {year} {2018})}\BibitemShut {NoStop}%
\bibitem [{\citenamefont {Mock}\ \emph {et~al.}(2017)\citenamefont {Mock}, \citenamefont {Korlacki}, \citenamefont {Knight},\ and\ \citenamefont {Schubert}}]{Mock2017}%
  \BibitemOpen
  \bibfield  {author} {\bibinfo {author} {\bibfnamefont {A.}~\bibnamefont {Mock}}, \bibinfo {author} {\bibfnamefont {R.}~\bibnamefont {Korlacki}}, \bibinfo {author} {\bibfnamefont {S.}~\bibnamefont {Knight}}, \ and\ \bibinfo {author} {\bibfnamefont {M.}~\bibnamefont {Schubert}},\ }\href {\doibase 10.1103/PhysRevB.95.165202} {\bibfield  {journal} {\bibinfo  {journal} {Phys. Rev. B}\ }\textbf {\bibinfo {volume} {95}},\ \bibinfo {pages} {165202} (\bibinfo {year} {2017})}\BibitemShut {NoStop}%
\bibitem [{\citenamefont {Stokey}\ \emph {et~al.}(2020)\citenamefont {Stokey}, \citenamefont {Mock}, \citenamefont {Korlacki}, \citenamefont {Knight}, \citenamefont {Darakchieva}, \citenamefont {Schöche},\ and\ \citenamefont {Schubert}}]{Stokey}%
  \BibitemOpen
  \bibfield  {author} {\bibinfo {author} {\bibfnamefont {M.}~\bibnamefont {Stokey}}, \bibinfo {author} {\bibfnamefont {A.}~\bibnamefont {Mock}}, \bibinfo {author} {\bibfnamefont {R.}~\bibnamefont {Korlacki}}, \bibinfo {author} {\bibfnamefont {S.}~\bibnamefont {Knight}}, \bibinfo {author} {\bibfnamefont {V.}~\bibnamefont {Darakchieva}}, \bibinfo {author} {\bibfnamefont {S.}~\bibnamefont {Schöche}}, \ and\ \bibinfo {author} {\bibfnamefont {M.}~\bibnamefont {Schubert}},\ }\href {\doibase 10.1063/1.5135016} {\bibfield  {journal} {\bibinfo  {journal} {Journal of Applied Physics}\ }\textbf {\bibinfo {volume} {127}},\ \bibinfo {pages} {115702} (\bibinfo {year} {2020})},\ \Eprint {http://arxiv.org/abs/https://pubs.aip.org/aip/jap/article-pdf/doi/10.1063/1.5135016/13761410/115702\_1\_online.pdf} {https://pubs.aip.org/aip/jap/article-pdf/doi/10.1063/1.5135016/13761410/115702\_1\_online.pdf} \BibitemShut {NoStop}%
\bibitem [{\citenamefont {Richter}\ \emph {et~al.}(2024)\citenamefont {Richter}, \citenamefont {Knight}, \citenamefont {Bulancea-Lindvall}, \citenamefont {Mu}, \citenamefont {Kuehne}, \citenamefont {Stokey}, \citenamefont {Ruder}, \citenamefont {Rindert}, \citenamefont {Ivady}, \citenamefont {Abrikosov}, \citenamefont {de~Walle}, \citenamefont {Schubert},\ and\ \citenamefont {Darakchieva}}]{RichterPRBFebGO2024}%
  \BibitemOpen
  \bibfield  {author} {\bibinfo {author} {\bibfnamefont {S.}~\bibnamefont {Richter}}, \bibinfo {author} {\bibfnamefont {S.}~\bibnamefont {Knight}}, \bibinfo {author} {\bibfnamefont {O.}~\bibnamefont {Bulancea-Lindvall}}, \bibinfo {author} {\bibfnamefont {S.}~\bibnamefont {Mu}}, \bibinfo {author} {\bibfnamefont {P.}~\bibnamefont {Kuehne}}, \bibinfo {author} {\bibfnamefont {M.}~\bibnamefont {Stokey}}, \bibinfo {author} {\bibfnamefont {A.}~\bibnamefont {Ruder}}, \bibinfo {author} {\bibfnamefont {V.}~\bibnamefont {Rindert}}, \bibinfo {author} {\bibfnamefont {V.}~\bibnamefont {Ivady}}, \bibinfo {author} {\bibfnamefont {I.}~\bibnamefont {Abrikosov}}, \bibinfo {author} {\bibfnamefont {C.~V.}\ \bibnamefont {de~Walle}}, \bibinfo {author} {\bibfnamefont {M.}~\bibnamefont {Schubert}}, \ and\ \bibinfo {author} {\bibfnamefont {V.}~\bibnamefont {Darakchieva}},\ }\href@noop {} {\bibfield  {journal} {\bibinfo  {journal} {Phys. Rev. B}\ }\textbf {\bibinfo {volume} {XXX}},\ \bibinfo {pages} {X} (\bibinfo {year}
  {2024})}\BibitemShut {NoStop}%
\bibitem [{\citenamefont {Rindert}\ \emph {et~al.}(2024)\citenamefont {Rindert}, \citenamefont {Richter}, \citenamefont {Kühne}, \citenamefont {Ruder}, \citenamefont {Darakchieva},\ and\ \citenamefont {Schubert}}]{rindert2024bloch}%
  \BibitemOpen
  \bibfield  {author} {\bibinfo {author} {\bibfnamefont {V.}~\bibnamefont {Rindert}}, \bibinfo {author} {\bibfnamefont {S.}~\bibnamefont {Richter}}, \bibinfo {author} {\bibfnamefont {P.}~\bibnamefont {Kühne}}, \bibinfo {author} {\bibfnamefont {A.}~\bibnamefont {Ruder}}, \bibinfo {author} {\bibfnamefont {V.}~\bibnamefont {Darakchieva}}, \ and\ \bibinfo {author} {\bibfnamefont {M.}~\bibnamefont {Schubert}},\ }\href@noop {} {\enquote {\bibinfo {title} {Bloch equations in terahertz magnetic-resonance ellipsometry},}\ } (\bibinfo {year} {2024}),\ \Eprint {http://arxiv.org/abs/2404.12805} {arXiv:2404.12805 [cond-mat.mtrl-sci]} \BibitemShut {NoStop}%
\bibitem [{bol(2006)}]{boltonchapter10}%
  \BibitemOpen
  \enquote {\bibinfo {title} {Relaxation times, linewidths and spin kinetic phenomena},}\ in\ \href {\doibase https://doi.org/10.1002/9780470084984.ch10} {\emph {\bibinfo {booktitle} {Electron Paramagnetic Resonance}}}\ (\bibinfo  {publisher} {John Wiley \& Sons, Ltd},\ \bibinfo {year} {2006})\ Chap.~\bibinfo {chapter} {10}, pp.\ \bibinfo {pages} {301--356},\ \Eprint {http://arxiv.org/abs/https://onlinelibrary.wiley.com/doi/pdf/10.1002/9780470084984.ch10} {https://onlinelibrary.wiley.com/doi/pdf/10.1002/9780470084984.ch10} \BibitemShut {NoStop}%
\bibitem [{\citenamefont {Baranov}\ \emph {et~al.}(1997)\citenamefont {Baranov}, \citenamefont {Ilyin},\ and\ \citenamefont {Mokhov}}]{BARANOV1997611}%
  \BibitemOpen
  \bibfield  {author} {\bibinfo {author} {\bibfnamefont {P.}~\bibnamefont {Baranov}}, \bibinfo {author} {\bibfnamefont {I.}~\bibnamefont {Ilyin}}, \ and\ \bibinfo {author} {\bibfnamefont {E.}~\bibnamefont {Mokhov}},\ }\href {\doibase https://doi.org/10.1016/S0038-1098(96)00667-9} {\bibfield  {journal} {\bibinfo  {journal} {Solid State Communications}\ }\textbf {\bibinfo {volume} {101}},\ \bibinfo {pages} {611} (\bibinfo {year} {1997})}\BibitemShut {NoStop}%
\bibitem [{\citenamefont {Eaton}\ and\ \citenamefont {Eaton}(2000)}]{Eaton2000}%
  \BibitemOpen
  \bibfield  {author} {\bibinfo {author} {\bibfnamefont {S.~S.}\ \bibnamefont {Eaton}}\ and\ \bibinfo {author} {\bibfnamefont {G.~R.}\ \bibnamefont {Eaton}},\ }\enquote {\bibinfo {title} {Relaxation times of organic radicals and transition metal ions},}\ in\ \href {\doibase 10.1007/0-306-47109-4_2} {\emph {\bibinfo {booktitle} {Distance Measurements in Biological Systems by EPR}}},\ \bibinfo {editor} {edited by\ \bibinfo {editor} {\bibfnamefont {L.~J.}\ \bibnamefont {Berliner}}, \bibinfo {editor} {\bibfnamefont {G.~R.}\ \bibnamefont {Eaton}}, \ and\ \bibinfo {editor} {\bibfnamefont {S.~S.}\ \bibnamefont {Eaton}}}\ (\bibinfo  {publisher} {Springer US},\ \bibinfo {address} {Boston, MA},\ \bibinfo {year} {2000})\ pp.\ \bibinfo {pages} {29--154}\BibitemShut {NoStop}%
\bibitem [{\citenamefont {Abragam}\ and\ \citenamefont {Bleaney}(1970)}]{Abragam_Bleaney_1970}%
  \BibitemOpen
  \bibfield  {author} {\bibinfo {author} {\bibfnamefont {A.}~\bibnamefont {Abragam}}\ and\ \bibinfo {author} {\bibfnamefont {B.}~\bibnamefont {Bleaney}},\ }\href@noop {} {\emph {\bibinfo {title} {Electron paramagnetic resonance of transition ions}}}\ (\bibinfo  {publisher} {Clarendon Press},\ \bibinfo {year} {1970})\BibitemShut {NoStop}%
\bibitem [{\citenamefont {Slichter}(1990)}]{Slichter1990}%
  \BibitemOpen
  \bibfield  {author} {\bibinfo {author} {\bibfnamefont {C.~P.}\ \bibnamefont {Slichter}},\ }\enquote {\bibinfo {title} {Basic theory},}\ in\ \href {\doibase 10.1007/978-3-662-09441-9_2} {\emph {\bibinfo {booktitle} {Principles of Magnetic Resonance}}}\ (\bibinfo  {publisher} {Springer Berlin Heidelberg},\ \bibinfo {address} {Berlin, Heidelberg},\ \bibinfo {year} {1990})\ pp.\ \bibinfo {pages} {11--64}\BibitemShut {NoStop}%
\bibitem [{\citenamefont {Polder}(1949)}]{doi:10.1080/14786444908561215}%
  \BibitemOpen
  \bibfield  {author} {\bibinfo {author} {\bibfnamefont {D.}~\bibnamefont {Polder}},\ }\href {\doibase 10.1080/14786444908561215} {\bibfield  {journal} {\bibinfo  {journal} {The London, Edinburgh, and Dublin Philosophical Magazine and Journal of Science}\ }\textbf {\bibinfo {volume} {40}},\ \bibinfo {pages} {99} (\bibinfo {year} {1949})}\BibitemShut {NoStop}%
\bibitem [{\citenamefont {Lehmeyer}\ and\ \citenamefont {Merten}(1985)}]{magnonpolaritons_restrahlen}%
  \BibitemOpen
  \bibfield  {author} {\bibinfo {author} {\bibfnamefont {A.}~\bibnamefont {Lehmeyer}}\ and\ \bibinfo {author} {\bibfnamefont {L.}~\bibnamefont {Merten}},\ }\href {\doibase https://doi.org/10.1016/0304-8853(85)90083-6} {\bibfield  {journal} {\bibinfo  {journal} {Journal of Magnetism and Magnetic Materials}\ }\textbf {\bibinfo {volume} {50}},\ \bibinfo {pages} {32} (\bibinfo {year} {1985})}\BibitemShut {NoStop}%
\bibitem [{\citenamefont {Jensen}\ \emph {et~al.}(1997)\citenamefont {Jensen}, \citenamefont {Feiven}, \citenamefont {Parker},\ and\ \citenamefont {Camley}}]{Magneticpolaritonmodes}%
  \BibitemOpen
  \bibfield  {author} {\bibinfo {author} {\bibfnamefont {M.~R.~F.}\ \bibnamefont {Jensen}}, \bibinfo {author} {\bibfnamefont {S.~A.}\ \bibnamefont {Feiven}}, \bibinfo {author} {\bibfnamefont {T.~J.}\ \bibnamefont {Parker}}, \ and\ \bibinfo {author} {\bibfnamefont {R.~E.}\ \bibnamefont {Camley}},\ }\href {\doibase 10.1088/0953-8984/9/34/013} {\bibfield  {journal} {\bibinfo  {journal} {Journal of Physics: Condensed Matter}\ }\textbf {\bibinfo {volume} {9}},\ \bibinfo {pages} {7233} (\bibinfo {year} {1997})}\BibitemShut {NoStop}%
\bibitem [{\citenamefont {Kullmann}\ \emph {et~al.}(1984)\citenamefont {Kullmann}, \citenamefont {Strobel},\ and\ \citenamefont {Geick}}]{Kullmann_1984}%
  \BibitemOpen
  \bibfield  {author} {\bibinfo {author} {\bibfnamefont {W.}~\bibnamefont {Kullmann}}, \bibinfo {author} {\bibfnamefont {K.}~\bibnamefont {Strobel}}, \ and\ \bibinfo {author} {\bibfnamefont {R.}~\bibnamefont {Geick}},\ }\href {\doibase 10.1088/0022-3719/17/36/036} {\bibfield  {journal} {\bibinfo  {journal} {Journal of Physics C: Solid State Physics}\ }\textbf {\bibinfo {volume} {17}},\ \bibinfo {pages} {6855} (\bibinfo {year} {1984})}\BibitemShut {NoStop}%
\bibitem [{\citenamefont {Paskova}\ \emph {et~al.}(2006)\citenamefont {Paskova}, \citenamefont {Kroeger}, \citenamefont {Figge}, \citenamefont {Hommel}, \citenamefont {Darakchieva}, \citenamefont {Monemar}, \citenamefont {Preble}, \citenamefont {Hanser}, \citenamefont {Williams},\ and\ \citenamefont {Tutor}}]{Paskova_APL2006}%
  \BibitemOpen
  \bibfield  {author} {\bibinfo {author} {\bibfnamefont {T.}~\bibnamefont {Paskova}}, \bibinfo {author} {\bibfnamefont {R.}~\bibnamefont {Kroeger}}, \bibinfo {author} {\bibfnamefont {S.}~\bibnamefont {Figge}}, \bibinfo {author} {\bibfnamefont {D.}~\bibnamefont {Hommel}}, \bibinfo {author} {\bibfnamefont {V.}~\bibnamefont {Darakchieva}}, \bibinfo {author} {\bibfnamefont {B.}~\bibnamefont {Monemar}}, \bibinfo {author} {\bibfnamefont {E.}~\bibnamefont {Preble}}, \bibinfo {author} {\bibfnamefont {A.}~\bibnamefont {Hanser}}, \bibinfo {author} {\bibfnamefont {N.~M.}\ \bibnamefont {Williams}}, \ and\ \bibinfo {author} {\bibfnamefont {M.}~\bibnamefont {Tutor}},\ }\href {\doibase 10.1063/1.2236901} {\bibfield  {journal} {\bibinfo  {journal} {Applied Physics Letters}\ }\textbf {\bibinfo {volume} {89}},\ \bibinfo {pages} {051914} (\bibinfo {year} {2006})},\ \Eprint {http://arxiv.org/abs/https://pubs.aip.org/aip/apl/article-pdf/doi/10.1063/1.2236901/14351913/051914\_1\_online.pdf}
  {https://pubs.aip.org/aip/apl/article-pdf/doi/10.1063/1.2236901/14351913/051914\_1\_online.pdf} \BibitemShut {NoStop}%
\bibitem [{\citenamefont {Kühne}\ \emph {et~al.}(2018)\citenamefont {Kühne}, \citenamefont {Armakavicius}, \citenamefont {Stanishev}, \citenamefont {Herzinger}, \citenamefont {Schubert},\ and\ \citenamefont {Darakchieva}}]{2018LuEllipsometer}%
  \BibitemOpen
  \bibfield  {author} {\bibinfo {author} {\bibfnamefont {P.}~\bibnamefont {Kühne}}, \bibinfo {author} {\bibfnamefont {N.}~\bibnamefont {Armakavicius}}, \bibinfo {author} {\bibfnamefont {V.}~\bibnamefont {Stanishev}}, \bibinfo {author} {\bibfnamefont {C.~M.}\ \bibnamefont {Herzinger}}, \bibinfo {author} {\bibfnamefont {M.}~\bibnamefont {Schubert}}, \ and\ \bibinfo {author} {\bibfnamefont {V.}~\bibnamefont {Darakchieva}},\ }\href {\doibase 10.1109/TTHZ.2018.2814347} {\bibfield  {journal} {\bibinfo  {journal} {IEEE Transactions on Terahertz Science and Technology}\ }\textbf {\bibinfo {volume} {8}},\ \bibinfo {pages} {257} (\bibinfo {year} {2018})}\BibitemShut {NoStop}%
\bibitem [{\citenamefont {Huang}(1951)}]{1951Huang}%
  \BibitemOpen
  \bibfield  {author} {\bibinfo {author} {\bibfnamefont {K.}~\bibnamefont {Huang}},\ }\href {https://doi.org/10.1098/rspa.1951.0166} {\bibfield  {journal} {\bibinfo  {journal} {Proceedings of the Royal Society of London. Series A, Mathematical and Physical Sciences}\ }\textbf {\bibinfo {volume} {208}},\ \bibinfo {pages} {352} (\bibinfo {year} {1951})}\BibitemShut {NoStop}%
\bibitem [{\citenamefont {Pekar}(1958)}]{PEKAR195811}%
  \BibitemOpen
  \bibfield  {author} {\bibinfo {author} {\bibfnamefont {S.}~\bibnamefont {Pekar}},\ }\href {\doibase https://doi.org/10.1016/0022-3697(58)90127-6} {\bibfield  {journal} {\bibinfo  {journal} {Journal of Physics and Chemistry of Solids}\ }\textbf {\bibinfo {volume} {5}},\ \bibinfo {pages} {11} (\bibinfo {year} {1958})}\BibitemShut {NoStop}%
\bibitem [{\citenamefont {Hopfield}(1958)}]{PhysRev.112.1555}%
  \BibitemOpen
  \bibfield  {author} {\bibinfo {author} {\bibfnamefont {J.~J.}\ \bibnamefont {Hopfield}},\ }\href {\doibase 10.1103/PhysRev.112.1555} {\bibfield  {journal} {\bibinfo  {journal} {Phys. Rev.}\ }\textbf {\bibinfo {volume} {112}},\ \bibinfo {pages} {1555} (\bibinfo {year} {1958})}\BibitemShut {NoStop}%
\end{thebibliography}%
\end{document}